\input harvmac
\def\npb#1(#2)#3{{ Nucl. Phys. }{B#1} (#2) #3}
\def\plb#1(#2)#3{{ Phys. Lett. }{#1B} (#2) #3}
\def\pla#1(#2)#3{{ Phys. Lett. }{#1A} (#2) #3}
\def\prl#1(#2)#3{{ Phys. Rev. Lett. }{#1} (#2) #3}
\def\mpla#1(#2)#3{{ Mod. Phys. Lett. }{A#1} (#2) #3}
\def\ijmpa#1(#2)#3{{ Int. J. Mod. Phys. }{A#1} (#2) #3}
\def\cmp#1(#2)#3{{ Commun. Math. Phys. }{#1} (#2) #3}
\def\cqg#1(#2)#3{{ Class. Quantum Grav. }{#1} (#2) #3}
\def\jmp#1(#2)#3{{ J. Math. Phys. }{#1} (#2) #3}
\def\anp#1(#2)#3{{ Ann. Phys. }{#1} (#2) #3}
\def\prd#1(#2)#3{{ Phys. Rev.} {D\bf{#1}} (#2) #3}

\def\inbar{\,\vrule height1.5ex width.4pt depth0pt}
\def\IQ{\relax\,\hbox{$\inbar\kern-.3em{\rm Q}$}}
\def\IB{\relax{\rm I\kern-.18em B}}
\def\IC{\relax\hbox{$\inbar\kern-.3em{\rm C}$}}
\def\IP{\relax{\rm I\kern-.18em P}}
\def\IR{\relax{\rm I\kern-.18em R}}
\def\ZZ{\relax\ifmmode\mathchoice
{\hbox{Z\kern-.4em Z}}{\hbox{Z\kern-.4em Z}}
{\lower.9pt\hbox{Z\kern-.4em Z}}
{\lower1.2pt\hbox{Z\kern-.4em Z}}\else{Z\kern-.4em Z}\fi}

\def\n*#1{\nu^{* (#1)}}

\def\IP{{\bf P}}

\def\n#1{\nu_{#1}^*}

\def\IP{{\bf P}}

\def\n#1{\nu_{#1}^*}

\def\({ \left(  }
\def\){ \right) }

\def\-{\phantom{-}}
\noblackbox

\catcode`@=12
 \baselineskip16pt
 \noblackbox
\newif\ifdraft

\noblackbox
\newif\ifhypertex
\ifx\hyperdef\UnDeFiNeD
    \hypertexfalse
    \message{[HYPERTEX MODE OFF}
    \def\hyperref#1#2#3#4{#4}
    \def\hyperdef#1#2#3#4{#4}
    \def\hypernoname{}
    \def\e@tf@ur#1{}
    \def\hth/#1#2#3#4#5#6#7{{\tt hep-th/#1#2#3#4#5#6#7}}
    
\else
    \hypertextrue
    \message{[HYPERTEX MODE ON}
\def\hth/#1#2#3#4#5#6#7{
  {\tt hep-th/#1#2#3#4#5#6#7}}

\fi

\catcode`\@=11
\newif\iffigureexists
\newif\ifepsfloaded
\def\epsfcheck{
\ifdraft
\input epsf\epsfloadedtrue
\else
  \openin 1 epsf
  \ifeof 1 \epsfloadedfalse \else \epsfloadedtrue \fi
  \closein 1
  \ifepsfloaded
    \input epsf
  \else
\immediate\write20{NO EPSF FILE --- FIGURES WILL BE IGNORED}
  \fi
\fi
\def\epsfcheck{}}
\def\checkex#1{
\ifdraft
\figureexistsfalse\immediate%
\write20{Draftmode: figure #1 not included}
\else\relax
    \ifepsfloaded \openin 1 #1
        \ifeof 1
           \figureexistsfalse
  \immediate\write20{FIGURE FILE #1 NOT FOUND}
        \else \figureexiststrue
        \fi \closein 1
    \else \figureexistsfalse
    \fi
\fi}
\def\missbox#1#2{$\vcenter{\hrule
\hbox{\vrule height#1\kern1.truein
\raise.5truein\hbox{#2} \kern1.truein \vrule} \hrule}$}
\def\lfig#1{
\let\labelflag=#1%
\def\numb@rone{#1}%
\ifx\labelflag\UnDeFiNeD%
{\xdef#1{\the\figno}%
\writedef{#1\leftbracket{\the\figno}}%
\global\advance\figno by1%
}\fi{\hyperref{}{figure}{{\numb@rone}}{Fig.{\numb@rone}}}}
\def\figinsert#1#2#3#4{
\epsfcheck\checkex{#4}%
\def\figsize{#3}%
\let\flag=#1\ifx\flag\UnDeFiNeD
{\xdef#1{\the\figno}%
\writedef{#1\leftbracket{\the\figno}}%
\global\advance\figno by1%
}\fi
\goodbreak\midinsert%
\iffigureexists
\centerline{\epsfysize\figsize\epsfbox{#4}}%
\else%
\vskip.05truein
  \ifepsfloaded
  \ifdraft
  \centerline{\missbox\figsize{Draftmode: #4 not included}}%
  \else
  \centerline{\missbox\figsize{#4 not found}}
  \fi
  \else
  \centerline{\missbox\figsize{epsf.tex not found}}
  \fi
\vskip.05truein
\fi%
{\smallskip%
\leftskip 4pc \rightskip 4pc%
\noindent\ninepoint\sl \baselineskip=11pt%
{\bf{\hyperdef\hypernoname{figure}{{#1}}{Fig.{#1}}}:~}#2%
\smallskip}\bigskip\endinsert%
}
%


\nopagenumbers\abstractfont\hsize=\hstitle
\null
\rightline{\vbox{\baselineskip12pt\hbox{CALT-68-2225}
                                  \hbox{NSF-ITP-99-32}
                                  \hbox{hep-th/9905097}}}%
\vfill
\centerline{\titlefont  The AdS/CFT correspondence and  }
\vskip5pt
\centerline{\titlefont  Spectrum Generating Algebras}
\abstractfont\vfill\pageno=0

\vskip-1.0cm
\centerline{P. Berglund$^1$, E. G. Gimon$^2$ and
D. Minic$^{3}$\footnote{$^{}$}
      {Email: berglund@itp.ucsb.edu, egimon@theory.caltech.edu,
minic@physics.usc.edu }}
 \vskip .20in
 \centerline{\it $^1$Institute for Theoretical Physics}          \vskip-.4ex
 \centerline{\it University of California}         \vskip-.4ex
 \centerline{\it Santa Barbara, CA 93106, USA}       \vskip-.0ex
\vskip .05in
 \centerline{\it $^2$California Institute for Technology} \vskip-.4ex
 \centerline{\it Pasadena, CA 91125, USA}       \vskip-.0ex
\vskip .05in
 \centerline{\it $^3$Department of Physics and Astronomy} \vskip-.4ex
 \centerline{\it University of Southern California}         \vskip-.4ex
 \centerline{\it Los Angeles, CA 90089-0484, USA }       \vskip-.0ex
\vskip-.4ex
\vfill
\vskip-0.3cm
\vbox{\narrower\baselineskip=12pt\noindent
    We list the spectrum generating algebras for string theory
and M-theory compactified on various backgrounds of the form
$AdS_{d+1} \times S^n$.  We identify the representations of these
algebras which make up the classical supergravity spectra and
argue for the presence of these spectrum generating algebras in
the classical string/M-theory.  We also discuss the role of the
spectrum generating algebras on the conformal field theory side. }

\Date{}

\vfill\eject
\baselineskip=14pt plus 1 pt minus 1 pt

\lref\maldacena{J. Maldacena, Adv. Theor. Math. Phys. {\bf 2} (1998) 231.}
 \lref\gubser{S.S. Gubser, I. R. Klebanov and A. M. Polyakov, Phys.
Lett. {\bf B428} (1998) 105; E. Witten, Adv. Theor. Math. Phys.
{\bf 2} (1998) 253.}
 \lref\duff{M. Duff, B. E. Nilsson and C. Pope,  Phys. Rept. {\bf 130}
 (1986) 1.}
 \lref\grw{M. G\"{u}naydin, L. J. Romans, N. P. Warner,
Phys. Lett. {\bf B146} (1984) 401.}
 \lref\sga{A. Barut, A. B\"{o}hm and Y. Ne'eman, Dynamical Groups and
Spectrum Generating Algebras, (World Scientific, 1988).}
 \lref\salamone{A. Salam and J. Strathdee, Ann. of Phys. {\bf 141}
(1982) 316. }
 \lref\salamtwo{A. Salam and E. Sezgin, Supergravity in Different
Dimensions (World Scientific, Singapore, 1989).}
 \lref\vilenkin{N. Ja. Vilenkin and A. U. Klimyk, Representation of
Lie Groups and Special Functions, vol. 1,2,3 (Kluwer Academic
Publishing, 1993).}
 \lref\fritz{F. Schwarz, Jour. Math. Phys. {\bf 12} (1971) 131.}
 \lref\adsfour{E. Sezgin, Phys. Lett. {\bf B138} (1984) 57;
Fortsch. Phys. {\bf 34} (1986) 217}
 \lref\adsfive{H. J. Kim, L. J. Romans, P. van Nieuwenhuizen, Phys.
Rev. {\bf D32} (1985) 389; M. G\"{u}naydin and N. Marcus, Class.
Quantum Grav. {\bf 2} (1985) L11.}
 \lref\gm{M. G\"{u}naydin and D. Minic, Nucl.Phys. {\bf B523} (1998) 145.}
 \lref\adsseven{K. Pilch, P. K. Townsend and  P. van Nieuwenhuizen,
Nucl. Phys. {\bf B242} (1984) 377; M. G\"{u}naydin, P. van
Nieuwenhuizen and N. P. Warner, Nucl. Phys. {\bf B255} (1985) 63. }
 \lref\adstwo{A. Strominger, hep-th/9809027; J.~Maldacena, J.~Michelson and A.~Strominger,
 JHEP {\bf 02}, 011 (1999) hep-th/9812073.}
 \lref\ferrara{S. Ferrara and A. Zaffaroni, hep-th/9807090;
L. Andrianopoli and S. Ferrara, hep-th/9807150.}
 \lref\gmz{M. G\"{u}naydin, D. Minic and M. Zagermann, Nucl. Phys.
{\bf B534} (1998) 96; erratum - ibid {\bf B538} (1999) 531; also,
hep-th/9810226, to appear in Nucl. Phys. B.}
 \lref\kallosh{
 R. R. Metsaev and A. Tseytlin, Nucl. Phys. {\bf B533} (1998) 109;
 R. Kallosh and  A. Tseytlin, JHEP {\bf 9810} (1998) 016;
 R.~Kallosh, J.~Rahmfeld and A.~Rajaraman, JHEP {\bf 09}, 002 (1998) hep-th/9805217;
 R.~Kallosh and J.~Rahmfeld,  Phys. Lett. {\bf B443}, 143 (1998)
 hep-th/9808038;
 J.~Rahmfeld and A.~Rajaraman,  hep-th/9809164;
 A.~Rajaraman and M.~Rozali, hep-th/9902046;
 J.~de Boer and S.L.~Shatashvili, hep-th/9905032;
 J.~de Boer, H.~Ooguri, H.~Robins and J.~Tannenhauser, JHEP {\bf 12}, 026
 (1998) hep-th/9812046;
 D.~Berenstein and R.G.~Leigh,  hep-th/9904104;
 A.~Giveon, D.~Kutasov and N.~Seiberg,
 Adv. Theor. Math. Phys. {\bf 2}, 733 (1998) hep-th/9806194;
 S.~Elitzur, O.~Feinerman, A.~Giveon and D.~Tsabar,  Phys. Lett.
 {\bf B449}, 180 (1999) hep-th/9811245;
 D.~Kutasov and N.~Seiberg, JHEP {\bf 04}, 008 (1999) hep-th/9903219;
 I.~Pesando, hep-th/9903086;
 I.~Pesando, JHEP {\bf 02}, 007 (1999) hep-th/9809145;
 M.~Bershadsy, S.~Zhukov and A.~Vaintrob,  hep-th/9902180;
 J.~Park and S.~Rey, JHEP {\bf 11}, 008 (1998) hep-th/9810154;
 J.~Park and S.~Rey, JHEP {\bf 01}, 001 (1999) hep-th/9812062;
 M.~Yu and B.~Zhang, hep-th/9812216.}
 \lref\berkovits{N. Berkovits, C. Vafa and E. Witten, hep-th/9902098.}
 \lref\orbifolds{S. Kachru and E. Silverstein, hep-th/9802183;
M. Bershadsky, Z. Kakushadze and C. Vafa, hep-th/9803076.}
 \lref\bonus{K. Intriligator, hep-th/9811047.}  \lref\eggz{J. Ellis,
M.K. Gaillard, M. Gunaydin, B. Zumino, Nucl.Phys. {\bf B224}
(1983) 427.}
 \lref\qdef{A. Jevicki and S. Ramgoolam, hep-th/9902059.}
 \lref\deboer{J. de Boer, hep-th/9806104}
 \lref\ofer{O.~Aharony and E.~Witten,  JHEP {\bf 11}, 018 (1998)
hep-th/9807205.}

\newsec{Introduction}
    There exists convincing evidence \refs{\maldacena, \gubser}
 for a duality between string
theory or M-theory on $AdS_{d+1} \times S^n$ with $N$ units of
$n$-form flux through $S^n$ and a $d$-dimensional $SU(N)$ superconformal
field
theory on the boundary of $AdS_{d+1}$.  This conjecture exists
both in a weak form and in a strong form.  In the weak form the
space $AdS_{d+1} \times S^n$, with size proportional to $N$, is
taken to be quite large.  In this limit supergravity dominates and
captures all the physics of the dual large $N$ superconformal field
theory. In the strong form of the conjecture, string theory or
M-theory effects need to be taken into account to properly
describe the finite $N$ superconformal field theory.  Available
evidence for the AdS/CFT conjecture focuses mainly on the weak
form \refs{\maldacena, \gubser}, although some progress has been made
 towards understanding
the full stringy spectrum~\kallosh.

    In order to understand string theory (M-theory is more
problematic) on $AdS_{d+1} \times S^n$, the classical string
action needs to be quantized in this background.  This procedure
should produce the discrete spectrum of string states and their
masses along with rules for calculating their interactions.  In
this paper, we use an alternative approach to provide information
on the string spectrum.  We consider the eleven, ten and
six-dimensional supergravity limits of M/string theory, as well as
massive ten-dimensional stringy fields expanded in Kaluza-Klein
(KK) modes on $S^n$. Even though identifying the proper
independent string degrees of freedom using this method is
extremely difficult, we argue that one important qualitative
feature of the Kaluza-Klein reduction survives, namely the
presence of a so-called {\it spectrum generating algebra}
\refs{\sga,\salamone,\grw}.

    A spectrum generating algebra (SGA) typically does not commute
with the Hamiltonian and is non-linearly realized at the level of
the action, but it describes the entire spectrum of a particular
physical system~\sga. SGAs have been very successfully used in
nuclear, atomic and molecular physics, not only in the study of
spectra but also in the computation of various transition
amplitudes (for more details on this subject, consult \sga).

  In Kaluza-Klein reductions SGAs usually appear because the towers
of harmonics used in these reductions can be fit into unitary
irreducible representations (UIR) of the conformal groups of the
corresponding compactified spaces (spheres, products of spheres,
or any Einstein spaces which have a natural action of the
conformal group)~\salamone, \grw, \gm.   Since the eigenvalues of
harmonics are related to the masses of the corresponding
Kaluza-Klein states, the algebra of the conformal group does not
commute with the Hamiltonian.  For the case of compactifications
on $S^n$ the corresponding conformal group $SO(n+1,1)$ acts as a
spectrum generating algebra~\refs{\salamone,\grw}.

The conformal generators of the SGA are not isometries of the
compactification manifold. Rather, the operation of rescaling of
the manifold corresponds to the ``scanning'' of the spectrum of
the associated operator (Dirac, Laplace, etc.) on the manifold in
question. In particular the spherical harmonics on $S^{n}$ provide
a natural UIR of the conformal group of $S^n$~\foot{This follows
from an extension of the Peter-Weyl theorem, \vilenkin, vol.2,
chapter 9.}, which generates the spectrum of KK modes, see
section~2.  We will explicitly demonstrate this construction for
supergravity fields. To do this, we extend the results known in
the supergravity literature~\grw\ by demonstrating how the
corresponding spectra fit into UIRs of the relevant conformal
group for some of the maximally symmetric examples of the AdS/CFT
correspondence: IIB string theory on $AdS_5 \times S^5$ \adsfive,
M-theory on $AdS_4 \times S^7$ \adsfour\ and $AdS_7 \times S^4$
\adsseven, and IIA or IIB string theory  on $AdS_3 \times S^3
\times X^4$ \deboer (see section~3 and Tables 1.-4.).  Our results
can be also extended to the case of IIA or IIB string theory on
$AdS_2 \times S^{2} \times X^6$ \adstwo.\foot{We can as well
consider heterotic/M-theory/F-theory compactifications on
Calabi-Yau manifolds to six and four dimensions as our starting
point.}

In view of the AdS/CFT correspondence we consider the map between
the action of an SGA on the supergravity spectrum and the
corresponding action of what we call an {\it operator generating
algebra} (OGA) on the chiral primaries  on the CFT side.
Demonstrating the presence of SGAs in the supergravity spectrum
allows us to argue for and better understand the extension of the
SGAs to the full string/M-theory. More explicitly, we discuss the
KK towers of the level one massive string states of the flat
ten-dimensional (IIA or IIB) string on $S^{5}$, and show how they
provide UIRs of the corresponding SGA. We expect that the action
of the OGA generalizes to include operators in the CFT dual to
stringy fields. In particular we discuss how the relevant OGA
could act on the so-called Konishi supermultiplet \refs{\ferrara,
\gmz} which is expected to correspond to massive string states of
IIB string theory on $S^{5}$, see section~4. Finally, in
section~5, we discuss the relevance of SGA in the case of the
recent proposal on the finite $N$ case and quantum deformed
isometries~\qdef.

\newsec{Review of representation theory of $SO(n+1,1)$}

Let us consider a generic supergravity theory compactified on $AdS_{m}
 \times S^{n}$. All supergravity fields can be expanded into harmonic
 functions on $S^{n}$ (this is just the physical statement of the
  Peter-Weyl theorem \vilenkin).
It can be shown that these harmonic functions provide a UIR of
 the conformal group on $S^{n}$, which is $SO(n+1,1)$.
This can be seen as follows
  \refs{\vilenkin,\grw}:
Let $S^n$ denote a unit sphere $\sum x^{i} x^{i} =1$ and let $g
\in SO(n+1,1)$ \eqn\cc{ g = \left ( \matrix{ a^{i}_{j} & b^{i} \cr
c_{j}     & d \cr } \right) } The action of $g \in SO(n+1,1)$ on
$S^{n}$ is given by \eqn\gaction{ (gx)^{i} = (a^{i}_{j}x^{j} +
b^{i})(c_{k}x^{k} + d)^{-1} } and the action of $g^{-1} \in
SO(n+1,1)$ on complex functions $f: S^{n} \rightarrow \IC$ by
\eqn\ginv{ ((g^{-1})f)(x^{i}) = (c_{k}x^{k} +
d)^{\sigma}f((gx)^{i}) } where $\sigma \in \IC$ is called the
weight. Furthermore, let $L^{2}(S^{n})$ be the Hilbert space of
square integrable complex functions over $S^{n}$ with the natural
inner product \eqn\inprod{ (f_{1}, f_{2}) \equiv \int_{S^{n}}
{\bar{f}}_{1}(x) f_{2}(x) \sqrt{G} d^{n}x } where $\sqrt{G}
d^{n}x$ is the $SO(n+1,1)$ invariant measure on $S^{n}$. Then it
is easy to show that this inner product is preserved under the
action of $g^{-1} \in SO(n+1,1)$ on $f_{1}$, $f_{2}$ defined
above, provided that the weight $\sigma = -n/2 + i \rho$, where
$\rho$ is an arbitrary real number. Thus the space of harmonic
functions over $S^{n}$ provides a unitary irreducible
representation of $SO(n+1,1)$~\vilenkin.

Since KK modes of supergravity fields on $S^{n}$ are expected to
fit into UIRs of the $SO(n+1,1)$ SGA, we obviously need to use the
representation theory of non-compact $SO(n+1,1)$ groups to
understand the physical spectrum of KK modes. Their exists a
construction for the UIRs of the group $SO(n+1,1)$~\fritz\
completely analogous to the one for UIRs of its maximal compact
subgroup $SO(n+1)$. In the case of $SO(n+1)$ a unitary irreducible
representation is determined by a set of numbers $m_{ij},\, (1
\leq i  < j \leq n+1)$, all of which are integer or half-integer
simultaneously (there are important differences between $n+1=2p$
and $n+1=2p+1$). A vector in the representation space is denoted
by $|m_{ij}>$, where $m_{ij}$ provide a complete set of highest
weight labels (named Gel'fand-Zetlin (GZ) labels) which uniquely
determine an irreducible representation. The labels $m_{ij}$ obey
the following conditions
\eqn\mcond{\eqalign{ -m_{2k+1,1} &\leq m_{2k,1} \leq m_{2k+1,1}
\leq \ldots\leq m_{2k+1, k-1} \leq m_{2k,k} \leq m_{2k+1, k}\cr
|m_{2k,1}| &\leq m_{2k-1,1} \leq m_{2k,2} \leq \ldots\leq m_{2k,
k-1} \leq m_{2k-1,k-1} \leq m_{2k, k}~,} }
where $k=1,\ldots,p-1$ if $n+1\in 2\ZZ$ and $k=1,\ldots,p$ if
$n+1\in 2\ZZ+1$.

Based on this result it can be shown~\fritz\ that the UIRs of
$SO(n+1,1)$ (with important differences between $n+1=2p$ and
$n+1=2p+1$) are described by a set of $SO(n+1)$ GZ labels
$m_{ij}$, satisfying certain inequalities, along with a weight
$\sigma = -n/2 + i \rho$. One important property of the UIRs of
$SO(n+1,1)$ is that irreducible representation of $SO(n+1)$ occur
within with multiplicity one or not at all \fritz.

For example, in the case of $SO(2p, 1)$ their exist UIRs with
$SO(2p)$ content described by the following requirement on the GZ
labels
 \eqn\mtwop{ |m_{2p,1}| \leq m_{2p+1,1} \leq \ldots \leq
 m_{2p+1, p-1} \leq m_{2p, p} }
where $m_{2p+1, j} =0, 1/2,1,\ldots $ for $1 \leq j \leq p-1$ and
the weight $\sigma = -p + i \rho$, with $\rho
> 0$.  These UIRs are labelled $D(m_{2p+1,1} \ldots m_{2p+1,p-1}, i\rho)$.
The complete list of UIRs  of $SO(n+1,1)$ in this notation is given
in~\fritz. We follow this notation and the results of~\fritz\ in
the main body of the paper. Note that in the GZ-notation these
representations typically consist of a finite number of infinite
towers.

   GZ labels form a  particularly convenient basis for understanding
the harmonic analysis on Kaluza-Klein (KK)
supergravity~\refs{\duff,\salamtwo} on any coset space
$G/H$~\salamone. In particular, the well known $AdS_m \times S^n$
backgrounds of KK supergravity can be understood as coset spaces,
upon the Euclideanization of the relevant $AdS_m$ spaces - $AdS_m
\rightarrow S^{m}$. Then the spectrum of KK supergravity on $AdS_m
\times S^n$ can be obtained from the harmonic analysis on $G/H =
SO(m+1)/SO(m) \times SO(n+1)/SO(n)$. In this analysis~\salamone,
one fixes the $H$ representations which describe the content of
all supergravity fields, and then one expands these fields in
terms of only those representations of $G$ which contain the fixed
$H$ representations. The GZ, or highest weight labels, provide a
natural basis for the implementation of this procedure~\adsfour.

More precisely, let the GZ labels of a fixed $H$ representation be
denoted by $(\alpha_{1}, \alpha_{2},\ldots , \alpha_{r})(
\beta_{1}, \beta_{2},\ldots, \beta_{q-1})$,  where \eqn\gzalpha{
\alpha_{1} \geq \alpha_{2} \geq\ldots \geq|\alpha_r|; \quad
\beta_{1}\geq \beta_{2} \geq\ldots \geq \beta_{q-1} } and
analogously, denote the GZ labels of a $G$ representation by
$(\gamma_{1}, \gamma_{2},\ldots, \gamma_{r})( \delta_{1},
\delta_{2},\ldots, \delta_{q})$, where~\foot{Here, for
illustrational purposes, we have assumed that $m+1\in 2\ZZ+1$ such
that the rank does not change between $SO(m+1)$ and $SO(m)$.
Similarly, we take $n+1\in 2\ZZ$. This also shows how the bound on
the $\alpha_i,\ldots,\delta_i$ are different depending on the
parity of $m+1$ ($n+1$).} \eqn\gzgamma{ \gamma_{1} \geq \gamma_{2}
\geq\ldots \geq \gamma_{r}; \quad \delta_{1}\geq \delta_{2}
\geq\ldots \geq |\delta_{q}|~. } Then according to a theorem by
Gel'fand and Zetlin~\vilenkin\ (vol. 3), the above $H$
representation is contained in the decomposition of the above $G$
representation provided \eqn\gzdelta{\eqalign{ \gamma_{1} &\geq
\alpha_{1} \geq \gamma_{2} \geq\ldots \geq \gamma_{r} \geq
|\alpha_{r}|\cr \delta_{1} &\geq \beta_{1} \geq \delta_{2}
\geq\ldots \geq \beta_{q-1}\geq |\delta_{q}|~.} } This theorem
combined with the representation theory of $SO(n+1,1)$ can be used
to easily read off the corresponding UIRs of the relevant
$SO(n+1,1)$ spectrum generating algebra, given the field content
of a particular supergravity theory~\foot{For the reader's
convenience we also give the relation between the Dynkin basis
(commonly used in the supergravity literature) and the GZ basis.
For the case of $SO(2p)$ denote the GZ basis by a set of integers
$(l_{1},l_{2},\ldots l_{p})$; then, the Dynkin labels are give by
a set of integers $(a_{1},a_{2},\ldots ,a_{p})$ such that $a_{1} =
l_{1} - l_{2},\ldots ,a_{p-1}=l_{p-1} - l_{p},
a_{p}=l_{p-1}+l_{p}$. Similarly, for the case of $SO(2p+1)$, one
has $a_{1} = l_{1} - l_{2},\ldots ,a_{p-1}=l_{p-1} - l_{p},
a_{p}=2l_{p}$.}.

\newsec{Type IIB supergravity on $AdS_5\times S^5$}
Given these technical tools, we now turn to actual physical
applications.  We consider the case of IIB supergravity on
$AdS_5\times S^5$, since for this case the actual boundary CFT of
the proposed duality is precisely defined; it is ${\cal{N}}=4$
$SU(N)$ super Yang-Mills theory (SYM) in four-dimensions.
Although we discuss in detail other supergravity $AdS_{d+1} \times
S^{n}$ examples (see Tables 2. - 4.), we study the $AdS_5 \times
S^5$ case (Table 1.) when we discuss the action of the SGA on the
full string theory.

 The bosonic sector of the ten-dimensional IIB supergravity consists of the
following representations of the little group $SO(8)$: \eqn\boson{
{\bf 1}_{\IC} + {\bf 28}_{\IC} + {\bf 35}_v + {\bf 35}_c~ } (the
dilaton and the axion, RR and NS 2-forms, the graviton, and the
self-dual RR 4-form).  The fermionic sector (spin 1/2 and spin 3/2
fields) is given by \eqn\fermion{ {\bf 8}_{\IC,s} + {\bf
56}_{\IC,s}~. } To understand the reduction of this spectrum on
$AdS_5\times S^5$, we first look at how the $SO(8)$ little group
representations break up into representations of $SO(5) \times
SO(3)$ on the {\it tangent bundle} of $AdS_5 \times S^5$. In
particular, we want to discuss the appearance of physical modes
(i.e., those modes that appear as poles in the $AdS_{5}$ bulk
propagators) and illustrate the general procedure by considering
only the bosonic fields \foot{We thank J. de Boer for very useful
discussions on this topic.}.

    On the tangent bundle of $AdS_5 \times S^5$ the ten-dimensional
little group $SO(8)$ splits into $SO(5) \times SO(3)$.  We start
our discussion by decomposing the $SO(8)$ representations for the
graviton, $h_{ab}$, and the self-dual four form, $a_{abcd}$, in
terms of $SO(5)\times SO(3)$~\foot{We use latin indices for
ten-dimensional fields, greek indices from the beginning of the
alphabet for $S^n$ and greek indices from the middle of the
alphabet for $AdS_{d+1}$.}. We get
 $$
 h_{ab}:\quad{\bf 35_v \to 1_1 + 1_5 + 5_3 + 14_1},
 $$
and
 $$ a_{abcd}:\quad{\bf 35_c \to 5_1 + 10_3},
 $$
respectively. We are interested in those representations of
$SO(6)\times SO(3)$ which contain the above representations of
$h_{ab}$ and $a_{abcd}$, since $SO(6)$ is the isometry group of
$S^5$. It is convenient to list these representations in terms of
their highest weight labels under $SO(6)$. The resulting $SO(6)$
labels, with their $SO(3)$ dimensions, are
 $$
 \eqalign{
&(l,0,0)_1,\quad (l,0,0)_5,\quad (l,0,0)_3,\,(l,1,0)_3,\quad\cr
&(l,0,0)_1,\,(l,1,0)_1,\,(l,2,0)_1}
 $$
for the graviton and
 $$ (l,0,0)_1,\,(l,1,0)_1,\quad
(l,1,0)_3,\,(l,1,\pm 1)_3
 $$
for the self-dual four form, respectively. In order to understand
which modes appear as physical from the point of view of the bulk
$AdS_5$ space we need to consider the action of the $AdS_5$ little
group $SO(4)$ on the above representations of $SO(6) \times
SO(3)$. These are uniquely lifted to representations of
$SO(6)\times SO(4)$, from which we directly read off the physical
modes propagating in the bulk $AdS_5$ space. We get
\eqn\physmodes{\eqalign{
&h_{\mu\nu}:\quad (l,0,0)_{(3,3)} \cr
&h_{\alpha\mu}:\quad (l,1,0)_{(2,2)}\cr
&h_{\alpha\beta}:\quad (l,2,0)_{(1,1)}\cr
&h^\alpha_\alpha:\quad (l,0,0)_{(1,1)}\cr
&a_{\alpha\beta\mu\nu}:\quad (l,1,1)_{(3,1)}+(l,1,-1)_{(1,3)}\cr
&a_{\alpha\beta\gamma\mu}:\quad (l,1,0)_{(2,2)}\cr
&a_{\alpha\beta\gamma\delta}:\quad (l,0,0)_{(1,1)}\cr}}
E.g., the ${\bf (3,3)}$ of $SO(4)$ is given in terms of ${\bf 1 +
3 + 5}$ of $SO(3)$ and so on. Note also that there will be a
mixing between modes with the same quantum numbers, such as
$h^\alpha_\alpha$ and $a_{\alpha\beta\gamma\delta}$. In the Tables
1.-4. we suppress this mixing and list the modes as above.

By comparing to \adsfive\ we see that group theory indeed accounts for all the
physical modes. One can also easily extend this analysis to
the fermionic part of the spectrum.

    The KK towers of physical modes cannot in general be fit alone
into UIRs of the conformal group of $S^{5}$ - $SO(6,1)$.  In order
to get full UIRs of $SO(6,1)$ we also need to consider gauge modes
(modes that do not appear as poles in the $AdS_5$ bulk
propagators).  The most convenient procedure for the
identification of KK towers of both physical and gauge modes, and
the corresponding UIRs of $SO(6,1)$, is to look at the Euclidean
$AdS_5 \times S^5$ space as a coset space - $G/H \equiv
SO(6)/SO(5) \times SO(6)/ SO(5)$.   We list the various KK modes
in terms of  $SO(5) \times SO(5)$ highest weight labels, and then
determine which $SO(6) \times SO(6)$ representations contain these
fixed $SO(5) \times SO(5)$ representations using the theorem of
Gel'fand and Zetlin reviewed in section 2. {}Here it is important
that we started with the full ten-dimensional tangent space
$SO(10)$ and not just the little group $SO(8)$ as we would
otherwise not see the gauge modes.  From the $SO(6)$ highest
weight, GZ labels, we directly read off the corresponding UIRs of
$SO(6,1)$. These UIRs must occur; the theorem above \vilenkin\
implies that a complete set of orthonormal harmonic functions on
$S^5$ forms a UIR of the conformal group of $S^5$, that is
$SO(6,1)$.

    To make our procedure described clearer, we choose
as an example the fields which come from the reduction of the
ten-dimensional graviton.  We write the $SO(5)$ representations of
these fields in terms of GZ labels; they are the $(0,0)_{GZ}$,
$(1,0)_{GZ}$ and $(2,0)_{GZ}$ representations.  From \gzdelta, the
$(2,0)_{GZ}$ representation of $SO(5)$, a scalar on $AdS_5$, is
contained in the $SO(6)$ representations with labels:
$(l+2,2,0)_{GZ}$, $(l+2,1,0)_{GZ}$, and $(l+2,0,0)_{GZ}$ ($l \geq
0$) which together form the $D^1(2; -5/2)$ UIR of SO(6,1).  Only
the symmetric tensors with $SO(6)$ labels $(l+2,2,0)_{GZ}$ are
physical, matching with $h_{\alpha\beta}$ in \physmodes, while the
others correspond to gauge modes. The $(1,0)_{GZ}$ representation
of $SO(5)$ is contained in the $SO(6)$ representations
$(l+1,1,0)_{GZ}$ (physical modes matching $h_{\alpha\mu}$) and
$(l+1,0,0)_{GZ}$ (gauge modes) which form the $D^1(1;-5/2)$
representation of $SO(6,1)$. Finally, the $(0,0)_{GZ}$
representation is contained in the $SO(6)$ representations
$(l,0,0)_{GZ}$ which form the $D^2(-5/2)$ representation of
$SO(6,1)$ (physical modes matching $h_{\mu\nu}$). The fields in
this tower couple to the symmetric trace operators on the CFT
side.

    In the discussion above, modes which usually are ignored because
they can be gauged away are {\it crucial} to the faithful action
of the conformal group of $S^5$ on the Kaluza-Klein spectrum.
Other gauge modes also appear in the spectrum in complete
representations of the SGA.  For example, an analysis of the mode
expansion on $AdS_5$ is enough to show that the ten-dimensional
graviton also yields a complete tower of vector gauge modes. We
will ignore these complete towers of gauge modes, and only mention
gauge modes which combine with physical modes to give UIRs of the
conformal group.  Generally, gauge modes are probably associated
with the diagonal $U(1)$ group on the boundary, whose role in the
$AdS/CFT$ duality is still not completely understood (see for
example \ofer).

    We now complete our analysis of the SGA representations which
appear in the supergravity spectrum of $AdS_5\times S^5$. The
antisymmetric tensor, the two-form $A_{\mu\nu}$, gives rise to a
tower $D^2(-5/2)$ of anti-symmetric chiral and anti-chiral tensor
fields, all describing physical modes. The vector $A_{\alpha\mu}$
gives rise to the towers of vectors that make up the $D^1(1;-5/2)$
representation with only the tower with modes of the form
$(l+1,1,0)_{GZ}$ in $D^1(1;-5/2)$ physical.  The scalar
$A_{\alpha\beta}$ gives rise to two physical KK-towers, modes of
the form $(l+1,1,\pm 1)_{GZ}$, which make up the $D^0(1,1;-5/2)$
representation.

    The rank-four antisymmetric self-dual tensor gives rise to
chiral and anti-chiral two-forms $A_{\alpha\beta\mu\nu}$ with
towers making up two $D^0(1,1;-5/2)$ representations. Each $D^0(1,1;-5/2)$
has two physical towers with $(l+1,1,\pm 1)_{GZ}$ of $SO(6)$,
adding up to four towers of physical two-forms . The vector
$A_{\alpha\beta\gamma\mu}$ gives a tower $D^1(1;-5/2)$ of vectors
but only the $(l+1,1,0)_{GZ}$ tower of $D^1(1;-5/2)$ describes
physical modes. The scalar mode, $A_{\alpha\beta\gamma\delta}$,
mixes with the $h^{\alpha}_{\alpha}$ scalar as can be seen from our earlier
discussion, with each of the mass eigenmodes giving rise to $D^2(-5/2)$.
Finally, the complex scalar, in terms of the axion and dilaton fields, gives
rise to yet one more physical KK-tower of complex scalars,
$D^2(-5/2)$.

The spin-1/2 field $\lambda$ gives twin towers of chiral and
anti-chiral spinors in the $D(1/2,1/2;-5/2+i\rho)$ representation.
Each of these contains physical modes $(l+1/2,1/2,\pm 1/2)_{GZ}$,
so each $D(1/2,1/2;-5/2)$ yields two towers. The chiral and
anti-chiral gravitini $\psi_\mu$ also come in the representation
$D(1/2,1/2;-5/2+i\rho)$, each with a total of two physical towers
with modes of the form $(l+1/2,1/2,\pm 1/2)_{GZ}$. Finally, we get
KK-towers of chiral and anti-chiral spin-$1/2$ fields from
$\psi_\alpha$, each in the $D(3/2,1/2;-5/2+i\rho)$ representation,
and each of these yielding physical modes of the form
$(l+3/2,1/2,\pm 1/2)_{GZ}$.
We summarize these results in Table 1. Tables 2. - 4. which contain
the fields and UIRs for the cases of $AdS_4\times S^7$, $AdS_7\times
S^4$ and $AdS_3\times S^3$ respectively, are obtained
following the same procedure.

\newsec{Conformal field theory, operator generating algebra,
massive string modes}

    We want to discuss what the SGA for the Kaluza-Klein states of
$AdS_{d+1}\times S^n$ means on the dual CFT side. We concentrate
on the $AdS_5/CFT_4$ correspondence, the ${\cal{N}}=4$ $SU(N)$
super Yang Mills theory in four-dimensions.  Other cases are more
difficult because the dual CFT is not easily described, though we
believe that similar arguments to those below can be applied there
as well.

We start from the fact that each supergravity KK tower corresponds
to a set of chiral primaries on the CFT side with appropriate
$SO(6)$ R-charges \gubser. Chiral primaries appear in the trace of
a symmetric product of ${\cal{N}}=4$ chiral superfields \ferrara.
For example, the traceless part of the following operator
 \eqn\chiralprime{
 Tr(W^{(i_{1}}W^{i_{2}}\cdot \ldots \cdot W^{i_{p})}) }
corresponds to the  KK states of IIB supergravity on $AdS_5 \times
S^5$, where $W$ is the ${\cal N}=4$ chiral superfield.  We have
shown that these KK states belong to UIRs of the $SO(6,1)$ SGA.
Given the map between KK modes and CFT chiral primaries, we
naturally expect that the complete set of UIRs of the $SO(6,1)$
SGA listed in Table~1 corresponds to
 \eqn\opuir{
 Tr(W^{i_{1}}) \oplus Tr(W^{(i_{1}}W^{i_{2})}) \oplus \ldots \oplus
 Tr(W^{(i_{1}}W^{i_{2}}\cdot \ldots \cdot W^{i_{p})})\oplus \ldots }
Note that there exists an ambiguity as to whether or not $W^i$
transforms in $SU(N)$ or $U(N)$ \gubser\ferrara . This ambiguity
is most likely related to the inclusion of gauge modes in the
complete $SO(6,1)$ UIRs. Taking the lowest component of \opuir,
the operators made up of the traceless part of
 \eqn\chiralprimephi{
 Tr(\phi^{(i_{1}}\phi^{i_{2}}\cdot \ldots \cdot \phi^{i_{p})}) }
($\phi$ is the $\theta^0\bar{\theta}^0$ component of the
${\cal{N}}=4$ chiral superfield $W$) fit into the $D^2(-5/2)$ of
representation of $SO(6,1)$.  Modulo subtleties involving gauge
modes and the extra $U(1)$, the other components of \chiralprime\
fill out the remaining UIRs listed in Table 1. This CFT
counterpart of the spectrum generating algebra of KK supergravity
we call operator generating algebra (OGA).

It is natural to ask whether this operator generating algebra
extends to all the operators in ${\cal{N}}=4$ SYM, including the
non-chiral ones which correspond to massive string modes. In order
to check this, we would have to classify and organize all the
non-chiral operators on the CFT side. We do not know of any such
classification. What we do know is that part of the $SO(6,1)$ OGA
acts on the chiral primaries by tensoring with a superfield in the
${\bf 6}$ of $SO(6)$ and symmetrizing.  How does this procedure
generalize to non-chiral primaries?  We sketch a natural proposal
as follows. Given an operator $Tr(O(W))$,  a set of UIRs of the
$SO(6,1)$ OGA is generated by the following operators
 \eqn\opuirguess{ Tr(O(W)W^{i_{1}}) \oplus Tr(O(W)W^{(i_{1}}W^{i_{2})})
   \oplus \ldots \oplus Tr(O(W)W^{(i_{1}}W^{i_{2}}\cdot \ldots \cdot
   W^{i_{p})})\oplus \ldots }
The fact that operators such as $Tr(O(W)W^{i_{1}})$ are not
necessarily irreducible and give direct sums of $SO(6)$
representations is useful for generating operators dual to both
physical modes and gauge modes.  Unfortunately, since we have not
explicitly determined the generators of the proposed $SO(6,1)$
OGA, we cannot actually prove that various operators belong to
UIRs of this OGA.  The issue also arises as to how to deal with
the possible mixing of different operators within the same UIR, a
problem which already exists for the chiral primary operators.

    Let us illustrate how our proposal for an OGA might work by
considering the so-called Konishi multiplet on the CFT side. In
terms of the SYM superfields, this multiplet is written as
$Tr(W_iW^i)$ \ferrara.  It has been suggested that the Konishi
multiplet corresponds to massive string states propagating in
$AdS_5$ \refs{\ferrara, \gmz}.  Consider the scalar operator in
the Konishi multiplet which is the $\theta^4\bar{\theta}^4$
component of $Tr(W_iW^i)$ and transforms in the ${\bf
105}=(4,0,0)_{GZ}$ of $SO(6)$.  Suppose we assume that it sits
naturally at the bottom of a $(l+4,0,0)_{GZ}$ KK tower of $SO(6)$.
On the $AdS$ side this is what we would expect from scalars coming
from a ten-dimensional four-tensor reduced on $S^5$. A good
candidate for the appropriate $SO(6,1)$ UIR is then
$D^{1}(4;-5/2)$.  It is made up of the towers
 \eqn\towers{ (l+4,0,0)_{GZ},\; (l+4,1,0)_{GZ},\; (l+4,2,0)_{GZ},\; (l+4,3,0)_{GZ},\;
 (l+4,4,0)_{GZ}.}
None of the extra towers in \towers\ have operators which can
appear in the Konishi multiplet, but if we take into account the
whole set of operators given by $Tr(W_iW^i W^{(i_1} \cdot \ldots
\cdot W^{i_p)})$, then at $p=1,2,3,4$ we find operators (dual to
bulk scalars) with $SO(6)$ weights
 \eqn\extras{(4,1,0)_{GZ},\; (4,2,0)_{GZ},\; (4,3,0)_{GZ},\;
(4,4,0)_{GZ}} which could sit at the bottoms of the extra towers.
It is important to note that we do not know whether the operators
above are dual to gauge modes or physical modes, since we lack a
precise rule for making this distinction.  Still, our primitive
fit for the Konishi multiplet is an indication that there might
exist an OGA, $SO(6,1)$, on the CFT side which organizes even the
non-chiral operators.

    Let us now address these issues from the $AdS$ side.
What happens with stringy, massive modes on the $AdS_5$ side?
These modes do not have protected anomalous dimensions on the CFT
side.  This is clear, since if we expand the stringy fields in
$S^5$ spherical harmonics their ten-dimensional masses will
contribute $\alpha '$ terms to their KK reduced AdS masses. The
non-linear nature of the equations relevant to stringy modes will
contribute further corrections and will also mix modes with the
same $SO(6)$ quantum numbers. Nevertheless, using the theorem
explained above \refs{\vilenkin, \grw}, the orthonormal basis of
harmonic functions on $S^5$ provides UIRs of the SGA $SO(6,1)$.
Thus, we expect that even the massive stringy modes can be fit in
UIRs of this SGA.  There is, however, a subtlety here: to {\it
prove} that $SO(6,1)$ is the SGA of the full IIB string theory on
$AdS_5 \times S^5$ we need to identify {\it all} the KK modes
generated by the massive stringy modes, and then fit them
explicitly (as we have done with the massless KK modes) in the
relevant UIRs of $SO(6,1)$.

One way of getting to the string theory on $AdS_5 \times S^5$ is
to start with the flat ten-dimensional string and then perturb it
with an RR operator (as in \berkovits) such that the theory flows
to the $AdS_5 \times S^5$ background \kallosh. One can contemplate
a connection between the large radius limit of $AdS_5 \times S^5$
and the flat ten-dimensional space, by taking $N \rightarrow
\infty$ and keeping $g_{YM}$ finite on the CFT side. In this limit
the states from the $AdS_5$ side should presumably map into states
propagating in the flat ten-dimensional space, the corresponding
vertex operators should match, etc.  The KK reduction of the
massive string modes on $S^5$ from flat ten dimensions should get
rearranged into the massive spectrum of the string on $AdS_5
\times S^5$. Also, on both sides there should exist a natural
action of the conformal group of $S^5$. If we can show that this
group acts as an SGA on the quantum ten-dimensional spectrum
reduced on $S^5$, we expect the conformal group of $S^5$ to appear
as an SGA for the string theory quantized directly about the
$AdS^5\times S^5$ background.

    Let us examine the KK reduction of the first massive level of the
flat IIB string on $S^5$. The multiplet transforms in the $({\bf
44 + 84 +128})^2$ of $SO(9)$ and has $256^2$ states. We consider
perturbations around the classical solution caused by the presence
of massive string modes in this multiplet. We apply the same
harmonic analysis used on the supergravity modes, and decompose
the $SO(9)$ representations coming from $({\bf 44 + 84 +128})^2$
in terms of $SO(5)\times SO(4)$ representations~\foot{Note that we
automatically get representations of the little group, $SO(4)$, of
$AdS_5$.}.  As before, we work in the GZ basis, which enables us
to read off the corresponding UIRs of $SO(6,1)$.  We consider
briefly one example of this particular procedure.  If we look at
${\bf 44 \times 44}$ we find a ${\bf 450}$ of $SO(9)$, in addition
to other representations of $SO(9)$ which we will ignore for now.
The ${\bf 450}$ is a four-tensor field in ten dimensions. It
decomposes into - among others - a $(4,0)_{GZ}$ of $SO(5)$ which
is contained in $(l+4,4,0)_{GZ}$, $(l+4,3,0)_{GZ}$,
$(l+4,2,0)_{GZ}$, $(l+4,1,0)_{GZ}$, $(l+4,0,0)_{GZ}$ of $SO(6)$
and generates the $D^{1}(4;-5/2)$ UIR of $SO(6,1)$.  This
particular UIR appeared in our discussion of the Konishi
multiplet. The same procedure can be extended to all fields at
this massive level, and to all massive levels.

In our analysis group theory has supplied us with details about
the spectrum. However, there are subtleties which can only be
addressed by examining the corrected classical equations of
motion; proper identification of physical and gauge modes as well
as the mixing of various KK modes. These phenomena happen already
at the massless level, so they are not surprising.  These
subtleties do not change the fact that physical and/or gauge modes
form UIRs of $SO(6,1)$! So, the conclusion seems to be that the
$SO(6,1)$ SGA from supergravity extends to the full string theory.
Of course, in order to prove this statement one would have to
examine all massive modes explicitly, and address the question of
mixing and identification of physical and gauge modes.

\newsec{Discussion}

To conclude, in this paper we have listed the spectrum generating
algebras for string theory and M-theory compactified on various
backgrounds of the form $AdS_{d+1} \times S^n$.  We have
identified the representations of these algebras which make up the
classical supergravity spectra and we argued for the existence of
these spectrum generating algebras in the classical
string/M-theory.  We also discussed the role of the spectrum
generating algebras on the conformal field theory side in the
framework of AdS/CFT correspondence.

One case we have not explicitly considered but which can be
analyzed in the same way is the $AdS_2 \times S^2$ background. The
corresponding boundary theory is some sort of conformal quantum
mechanics, which is not well understood \adstwo. Whatever that
boundary theory might be, there should exist an $SO(3,1)$ SGA
algebra on the supergravity/string side and a corresponding OGA on
the conformal quantum mechanics side.

Our methods should also apply to the case of string theory on $AdS
\times S^{n}/G$ \orbifolds, where $G$ is a discrete subgroup of
the isometry group of the sphere. It would be interesting to
understand the action of SGAs in this case.

One problem where we expect the concept of SGAs to have a
dynamical meaning is in the computation of correlation functions
within the framework of AdS/CFT duality.

Finally, an interesting question regarding SGAs concerns their
interpretation in the finite $N$ case of AdS/CFT duality (strong
conjecture).  Jevicki and Ramgoolam \qdef\ have proposed that
quantum deformed isometries should be relevant in this case. We
note that there exists an analog of Peter-Weyl theorem for the
case of $SU(2)_q$ - see~\vilenkin (vol. 3). The harmonic functions
for a $q$-deformed sphere can be also found in~\vilenkin (vol. 3).
It seems natural to expect that the harmonic functions over
$SO(n)_q$ fit into UIRs of $SO(n+1,1)_q$, thus generalizing our
previous results. In view of the proposal put forward in \qdef, we
expect the full string theory on $AdS_{d+1}\times S^m$ to exhibit
q-deformed SGAs.

\vskip.5cm {\bf Acknowledgements}: We would like to thank
P. Aschieri, I. Bars,
J. deBoer, G. Chalmers, D. Gross,
M. G\"{u}naydin, T. H\"{u}bsch, K. Pilch, J.
Polchinski and S. Ramgoolam for interesting discussions. One of
us (D.M.) would specially like to thank M. G\"{u}naydin for
illuminating discussions in the very early stages of this work.
The work of P. Berglund is supported in part by the Natural
Science Foundation under Grant No. PHY94-07194. The work of E.
Gimon is supported in part by the U. S. Department of Energy under
Grant no. DE-FG03-92ER40701. The work of D. Minic is supported in
part by the U.S. Departement of Energy under Grant no.
DE-FG03-84ER40168. P.B. would like to thank  Argonne, Caltech and
LBL, Berkeley for their hospitality while some of this work was
carried out. E.G. would
also like to thank the Harvard theory group for their hospitality
while this work was in progress.
D.M. would like to thank  ITP, Santa Barbara and
Caltech for providing stimulating environments for research.

\vfill\eject

\vfill
\def\ss{\scriptstyle}
\leftskip .0cm \rightskip .0cm \noindent
$$
\vbox{\offinterlineskip\tabskip=0pt\halign{
\strut
\vrule\hfil~$\ss{#}$~\hfil&\hfil~$\ss{#}$~\hfil&\hfil~$\ss{#}$~\hfil&
\hfil~$\ss{#}$~\hfil\vrule\cr
\noalign{\hrule}
h_{\mu\nu}&h_{\alpha\nu}&h_{\alpha\beta}&h^\alpha_{\,\,\,\alpha}\cr
D^2(-5/2)&D^1(1;-5/2)&D^1(2;-5/2)&D^2(-5/2)\cr
\noalign{\hrule}
A_{\mu\nu}&A_{\alpha\mu}&A_{\alpha\beta}&\cr
D^2(-5/2)&D^1(1;-5/2)&D^0(1,1;-5/2)&\cr
\noalign{\hrule}
a_{\alpha\beta\mu\nu}&a_{\alpha\beta\gamma\mu}&a_{\alpha\beta\gamma\delta}&
a+i\phi\cr
D^0(1,1;-5/2)&D^1(1;-5/2)&D^2(-5/2)&D^2(-5/2)\cr
\noalign{\hrule}
\psi_\mu&\psi_\alpha&\lambda&\cr
D(1/2,1/2;-5/2+i\rho)&D(1/2,1/2;-5/2+i\rho)&D(3/2,1/2;-5/2+i\rho)&\cr
\noalign{\hrule}}}
$$
\leftskip .5cm \rightskip .5cm
\noindent{\ninepoint  \baselineskip=8pt
 {{\bf Table 1:} The $AdS_5$ field content of IIB supergravity
 compactified on $AdS_5\times S^5$ \adsfive, organized in UIRs of $SO(6,1)$.
}} \leftskip .0cm \rightskip .0cm \noindent \vskip2.0cm

\leftskip .0cm \rightskip .0cm \noindent $$
\vbox{\offinterlineskip\tabskip=0pt\halign{ \strut
\vrule\hfil~$\ss{#}$~\hfil&\hfil~$\ss{#}$~\hfil&\hfil~$\ss{#}$~\hfil&
\hfil~$\ss{#}$~\hfil\vrule\cr \noalign{\hrule}
h_{\mu\nu}&h_{\alpha\nu}&h_{\alpha\beta}&
h^\alpha_{\,\,\,\alpha},\,h^\mu_{\,\,\,\mu}\cr
D^3(-7/2)&D^2(1;-7/2)&D^2(2;-7/2)&D^3(-7/2)\cr \noalign{\hrule}
C_{\alpha\mu\nu}&C_{\alpha\beta\mu}&C_{\alpha\beta\gamma}&\cr
D^2(1;-7/2)&D^1(1,1;-7/2)&D^0(1,1,1;-7/2)&\cr \noalign{\hrule}
\psi_\mu&\psi_\alpha&\lambda&\cr
D(1/2,1/2,1/2;-7/2+i\rho)&D(1/2,1/2,1/2;-7/2+i\rho)&
D(3/2,1/2,1/2;-7/2+i\rho)&\cr \noalign{\hrule}}} $$ \leftskip .5cm
\rightskip .5cm \noindent{\ninepoint  \baselineskip=8pt {{\bf
Table 2:}  The $AdS_4$ field content of eleven-dimensional
supergravity compactified on $AdS_4\times S^7$ \adsfour, organized
in UIRs of $SO(8,1)$ (matching \grw). }} \leftskip .0cm \rightskip
.0cm \noindent \vskip2.0cm

\leftskip .0cm \rightskip .0cm \noindent $$
\vbox{\offinterlineskip\tabskip=0pt\halign{ \strut
\vrule\hfil~$\ss{#}$~\hfil&\hfil~$\ss{#}$~\hfil&\hfil~$\ss{#}$~\hfil&
\hfil~$\ss{#}$~\hfil\vrule\cr \noalign{\hrule}
h_{\mu\nu}&h_{\alpha\nu}&h_{\alpha\beta}&h^\alpha_{\,\,\,\alpha}\cr
D^2(-2)&D^1(1;-2)&D^1(2;-2)&D^2(-2)\cr \noalign{\hrule}
C_{\mu\nu\rho}&C_{\alpha\mu\nu}&C_{\alpha\beta\mu}&C_{\alpha\beta\gamma}\cr
D^2(-2)&D^1(1;-2)&D^0(1,1;-2)&D^2(-2)\cr \noalign{\hrule}
\psi_\mu&\psi_\alpha&&\cr
D(3/2,1/2;-2+i\rho)&D(1/2,1/2;-2+i\rho&&\cr \noalign{\hrule}}} $$
\leftskip .5cm \rightskip .5cm \noindent{\ninepoint
\baselineskip=8pt {{\bf Table 3:}  The $AdS_7$ field content of
eleven-dimensional supergravity compactified on $AdS_7\times S^4$
\adsseven, organized in UIRs of $SO(5,1)$. }} \leftskip .0cm
\rightskip .0cm \noindent

\vfill\eject

\leftskip .0cm \rightskip .0cm \noindent $$
\vbox{\offinterlineskip\tabskip=0pt\halign{ \strut
\vrule\hfil~$\ss{#}$~\hfil&\hfil~$\ss{#}$~\hfil&\hfil~$\ss{#}$~\hfil&
\hfil~$\ss{#}$~\hfil\vrule\cr \noalign{\hrule}
h_{\mu\nu}&h_{\alpha\nu}&h_{\alpha\beta}&\cr
D^1(-3/2)&D^0(1;-3/2)&D^0(2;-3/2)&\cr \noalign{\hrule}
A_{\mu\nu}&A_\mu&A_\alpha&\phi\cr
D^0(1;-3/2)&D^1(-3/2)&D^0(1;-3/2)&D^1(-3/2)\cr \noalign{\hrule}
\psi_\mu&\psi_\alpha&\lambda&\cr
D(1/2;-3/2+i\rho)&D(3/2;-3/2+i\rho)&D(1/2;-3/2+i\rho)&\cr
\noalign{\hrule}}} $$ \leftskip .5cm \rightskip .5cm
\noindent{\ninepoint  \baselineskip=8pt {{\bf Table 4:} The
$AdS_3$ field content of six-dimensional supergravity compactified
on $AdS_3\times S^3$ \deboer, organized in UIRs of $SO(4,1)$. }}
\leftskip .0cm \rightskip .0cm \noindent

\vfill\eject

\listrefs \bye